\documentclass[twocolumn]{article}
\usepackage[ansinew]{inputenc}
\usepackage{latexsym}
\newcommand{\be}{\begin{equation}}
\newcommand{\ee}{\end{equation}}
\newcommand{\bear}{\begin{eqnarray}}
\newcommand{\ear}{\end{eqnarray}}
\newcommand{\ba}{\begin{eqnarray*}}
\newcommand{\ea}{\end{eqnarray*}}

\newcommand{\dt}{\mbox{\boldmath$:$}}

\newcommand{\no}{\noindent}

\newcommand{\mZ}{{\mathsf{Z}}}

\newcommand{\opsigma}{\mbox{\boldmath$\sigma$}}
\newcommand{\opmu}{\mbox{\boldmath$\mu$}}

\begin{document}

\title{Two-Dimensional Order and Disorder Thermofields}
\author{\small{ L. V. Belvedere$^\ast$ }\\
\small{\it{Instituto de F\'{\i}sica - Universidade Federal Fluminense}}\\
\small{\it{Av. Litor\^anea S/N, Boa Viagem, Niter\'oi, CEP 24210-340}}\\
\small{\it{Rio de Janeiro, Brasil}}\\
\small{$\ast$\,\it{belve@if.uff.br}}\\
}

\date{\today}

\maketitle

\begin{abstract}
{\small The main objective of this paper was to obtain the two-dimensional order and disorder thermal operators using the {\it Thermofield Bosonization} formalism. We show that the general property of the two-dimensional world according with the bosonized Fermi field at zero temperature can be constructed as a product of an order and a disorder variables which satisfy a dual field algebra holds at finite temperature. The general correlation functions of the  order and disorder thermofields are obtained. }
\end{abstract}

\hrulefill
\setcounter{equation}{0}

{\small

The operator realization for bosonization of 
fermions in $1 + 1$ dimensions at zero temperature ($T = 0$) corresponds to a mapping of a 
Fermi field algebra into a Bose field algebra. This algebraic isomorphism defines a one-to-one mapping of 
the corresponding Hilbert space of states. At $T \neq 0$, it is not evident that the operator bosonization 
survives at non zero temperatures. By one side, in the fermionic theory, a unitary operator depending on the Fermi-Dirac statistical weigh implements the transformation that promotes the Fermi field algebra into a Fermi thermofield algebra. On the other hand, in the bosonized version of the theory, another unitary operator depending on the
Bose-Einstein statistical weight implements the transformation that promotes the Bose field algebra
into a Bose thermofield algebra. In this way, the surviving of the operator bosonization at $T \neq 0$ dependes 
on the ``transmutation'' from Fermi-Dirac to Bose-Einstein statistics. Indeed, in an amazing way 
this statistical transmutation occurs, as shown in Ref. \cite{ABR}. The operator formulation for bosonization of massless fermions in $1 + 1$ dimensions, at finite, non-zero temperature $T$ is presented in Ref. \cite{ABR}. The 
{\it thermofield bosonization} has been achieved in the framework of the  real time formalism of Thermofield 
Dynamics. The well known Fermion-Boson correspondences in $1 + 1$ dimensions at zero temperature are shown to hold also at finite temperature \cite{ABR}.

In Ref. \cite{Marino} the two-dimensional Fermi field operator with generalized statistics at $T = 0$ is considered as 
a product of order and disorder variables. The main purpose of the present paper is to fill a gap in the literature, by providing the generalization for finite temperature of the two-dimensional order and disorder operators within the Thermofield dynamics approach \cite{TFD,U,Das,Haag,O,Mats}. We use the Thermofield Bosonization formalism, introduced in Ref. \cite{ABR}, to construct Fermi thermofields out of order and disorder thermal operators, which  satisfy an algebra analogous to the dual algebra of order and disorder variables in statistical mechanics, as it was first discussed by Kadanoff and Ceva \cite{KC}. This streamlines the presentation of the thermofield bosonization discussed  in Ref. \cite{ABR}.

Within the Thermofield Dynamics approach \cite{TFD,U,Das,Haag,O,Mats} a Quantum Field Theory at finite temperature is constructed by doubling the numbers of degrees of freedom. This is performed by introducing the ``tilde'' operators corresponding to each of the operators describing the system considered. This fictitious system is an identical copy of the original system under consideration, which entails a doubling of the Hilbert space of states. To begin with, let us consider the Fermi field doublet $\big (\psi (x) , \widetilde\psi (x) \big )$ of the two-dimensional massless Thirring model at $T = 0$  \cite{ABR,Th} which is defined by the Lagrangian density
$$
{\cal L}\,=\,i\,\bar\psi\,\gamma^\mu\,\partial_\mu\,\psi\,+\,\frac{g^2}{2}\,\big ( \bar\psi \gamma^\mu \psi \big)\,\big ( \bar\psi \gamma_\mu \psi \big )\,-
$$
\be
\Big (
-\,i\,\widetilde{\bar\psi}\,\gamma^\mu\,\partial_\mu\,\widetilde\psi\,+\,\frac{g^2}{2}\,\big ( \widetilde{\bar\psi} \gamma^\mu \widetilde\psi \big)\,\big (\widetilde{\bar\psi} \gamma_\mu \widetilde\psi \big )\,\Big )\,.
\ee

\no The {\it bosonized}  Fermi field doublet $\big (\psi (x) ,\widetilde\psi (x)\big )$ at zero temperature, which provides the operator solution of the quantum equations of motion, is constructed as a product of an order and a disorder operators \footnote{We have suppressed constant multiplicative factors and Klein factors that are present in the bosonized form of $\psi$ \cite{ABR}.},
\be
\psi (x)\, =\, f (\varepsilon )\,\opsigma (x)\,\opmu (x)\,,
\ee
\be
\widetilde{\psi} (x)\, =\, f ( \varepsilon )\,\widetilde{\opsigma} (x)\,\widetilde{\opmu}(x)\,,
\ee

\no where $f (\varepsilon )$ is an appropriate normalization factor and the order and disorder operators at zero temperature ($T = 0$) \cite{Marino} are given in terms of Wick-ordered exponentials
\be\label{order1}
\opsigma (x)\,=\,\dt\,e^{\textstyle\,i\,\,a\,\gamma^5\,\phi (x)}\,\dt\,,
\ee
\be
\opmu (x)\,=\,\dt\,e^{\textstyle\,i\,b\,\int_{x^1}^\infty\,
d\,z^1\,\partial_0 \phi (x^0 , z^1 )}\,\dt\,,
\ee

\no and the corresponding ``tilde'' operators,
\be
\widetilde{\opsigma} (x)\,=\,\dt\,e^{\textstyle\,-\,i\,\,a\,\gamma^5\,\widetilde{\phi} (x)}\,\dt\,,
\ee
\be\label{dis1}
\widetilde{\opmu} (x)\,=\,\dt\,e^{\textstyle\,-\,i\,b\,\int_{x^1}^\infty\,
d\,z^1\,\partial_0 \widetilde{\phi} (x^0 , z^1)}\,\dt\,.
\ee

\no The ``tilde conjugation'' is defined by ${\widetilde{c\phi}} = c^\ast \widetilde\phi$ and the fields $\phi (x)$ and $\widetilde{\phi} (x)$ are free massless scalar fields.  The scale dimension of $\psi$ ($\widetilde\psi$) is given by
\be
d_\psi = \frac{a^2 + b^2}{4 \pi}\,.
\ee

\no The massless Thirring model is obtained with 
\be
a\,=\,\frac{\beta}{2}\,\,,\,\,b\,=\,\frac{2 \pi}{\beta}\,,
\ee

\no where
\be
\beta^2\,=\,\frac{4 \pi}{1\,-\,\frac{g^2}{\pi}}\,.
\ee

\no The canonical free massless Fermi field operator is obtained with $g = 0$, $\beta^2 = 4 \pi$ ($a = b = \sqrt \pi$).

Since at $T = 0$ the order and disorder operators (\ref{order1})-(\ref{dis1}) are given in terms of Wick-ordered exponentials of a free massless scalar field,
\be
W (x)\,=\,\dt\,e^{\,i\,\lambda\,\phi (x)}\,\dt\,,
\ee

\no  in order to obtain the corresponding thermal operators we shall consider the Wick-ordered exponential of a free scalar field at finite temperature $W (x ; \beta )$. Following Ref. \cite{ABR} the thermal Wick exponential is obtained from the  exponential at $T = 0$ by performing the following transformation \cite{ABR},
\be \label{W}
W (x ; \beta ) = U^{- 1}[\theta  (\beta )]\,W (x)\,U [\theta (\beta )]\,,
\ee

\no where the operator $U [ \theta (\beta )]$ is defined in terms of the bosonic creation and annihilation
operators by,
$$
U (\theta) = e^{ - i {\cal Q} (\theta)} = 
$$
\be
e^{\,- \int_{- \infty}^{+ \infty} dp^1 \Big ( a (p^1) \widetilde a (p^1) - 
a^\dagger (p^1) \widetilde a^\dagger (p^1) \Big )
\theta (\vert p^1\vert ,\beta )}\,,
\ee

\no the Bogoliubov parameter $\theta (\vert p^1 \vert, \beta )$ is  implicitely defined by
\be
\sinh \theta (\vert p^1 \vert ; \beta )\,=\,
\frac{e^{\,-\,\beta \vert p^1 \vert /2}}
{\sqrt{1 - e^{\,-\beta \vert p^1 \vert }}}\,,
\ee
\be
\cosh \theta (\vert p^1 \vert ; \beta )\,=\,\frac{1}{\sqrt{1 - e^{\,-\beta 
\vert p^1 \vert}}}\,,
\ee

\no and the Bose-Einstein statistical weight is 
\be
N_B (\vert p^1 \vert ; \beta ) = \sinh^2 \theta (\vert p^1 \vert ; \beta )\,=\,
\frac{1}{e^{\beta \vert p^1 \vert} - 1}\,.
\ee

\no The transformed annihilation operators are given by,
$$
a (p^1 ; \beta ) = U ( - \theta  )\, a (p^1) \,U (\theta ) = 
$$
\be
a (p^1) 
\cosh \theta (\vert p^1 \vert ; \beta ) -
\widetilde a^\dagger (p^1) \sinh \theta (\vert p^1\vert ; \beta )\,,
\ee
$$
\widetilde a (p^1 ; \beta ) = U ( - \theta ) \widetilde a (p^1) U (\theta ) = 
$$
\be
\widetilde a (p^1) \cosh \theta (\vert p^1\vert ; \beta ) -
a^\dagger (p^1) \sinh \theta (\vert p^1\vert ; \beta )\,.
\ee

\no Using (\ref{W}) we obtain the order and disorder thermal operators $\{\opsigma (x ; \beta ), \opmu (x ; \beta ), \widetilde{\opsigma} (x ; \beta ), \widetilde{\opmu} (x ; \beta ) \}$, which are given in terms of the free massless scalar thermofields $\phi (x ; \beta )$ and $\widetilde\phi (x ; \beta )$,
\be\label{order}
\opsigma (x ; \beta )\,=\,[\mZ (\beta , \mu^\prime) ]^{\frac{1}{2}}\,\dt\,e^{\textstyle\,i\,a\,\gamma^5\,\phi (x ; \beta)}\,\dt\,,
\ee
\be\label{disorder}
\opmu (x ; \beta ) \,=\,[\mZ (\beta , \mu^\prime ]^{\frac{1}{2}}\,\dt\,e^{\textstyle\,i\,b\,\int_{x^1}^\infty\,
d\,z^1\,\partial_0 \phi (x^0 , z^1 ; \beta )}\,\dt\,,
\ee
\be\label{torder}
\widetilde{\opsigma} (x ; \beta)\,=\,[\mZ (\beta ; \mu^\prime )]^{\frac{1}{2}}\,\dt\,e^{\textstyle\,-\,i\,
a\,\gamma^5\,\widetilde\phi (x ; \beta)}\,\dt\,,
\ee
\be\label{tdisorder}
\widetilde{\opmu} (x ; \beta ) \,=\,[\mZ (\beta ; \mu^\prime )]^{\frac{1}{2}}\,\dt\,e^{\textstyle\,-\,
i\,b\,\int_{x^1}^\infty\,
d\,z^1\,\partial_0 \widetilde\phi (x^0 , z^1 ; \beta )}\,\dt\,.
\ee

\no The factor $\mZ (\beta , \mu^\prime )$ plays a role of a wave function renormalization for the thermal Wick-ordered exponentials \cite{ABR},
\be
\mZ ( \beta ; \mu^\prime )\,=\,e^{\textstyle\,-\,\frac{a^2 + b^2}{4 \pi}\,\mathit{z} (\beta ; \mu^\prime )}\,,
\ee

\no where $\mu^\prime$ is an infrared (IR) cut-off and the IR divergent factor $\mathit{z} (\beta ; \mu^\prime )$ is the mean number of particles having momenta $p$ in the range $[\mu^\prime , \infty )$,
\be
\mathit{z} ( \beta ; \mu^\prime )\,=\,\int_{\mu^\prime}^\infty\,\frac{dp}{p}\,N_B ( p ; \beta )\,.
\ee

\no As shown in Ref. \cite{ABR} the superselection rule associated with the thermal Wick exponential ensures the independence of physical quantities of the IR cut-off $\mu^\prime$.

As in the $T = 0$ case \cite{Marino}, the operators $\opmu (x^0 , y^1 ; \beta)$ 
and $\widetilde{\opmu} (x^0 , y^1 ; \beta )$ produce the following  transformations on the
thermal fields $\phi (x ; \beta)$ and $\widetilde\phi (x ; \beta)$, respectively,
\be
\opmu (x^0 , y^1 ; \beta )\,\,:\,\,\phi (x ; \beta ) \rightarrow  \phi (x ; \beta )\, +\, b\,\theta(y^1 - x^1)\,,
\ee
\be
\widetilde{\opmu} (x^0 , y^1 ; \beta )\,\,:\,\,\widetilde\phi (x ; \beta ) \rightarrow  \widetilde\phi (x ; \beta)\, +\,b\,\theta(y^1 - x^1)\,.
\ee

\no This transformation is a symmetry of the bosonized thermal Thirring model, which is described by a free massless scalar thermofield \cite{ABR}.  The thermal operators (\ref{order}-\ref{tdisorder}) satisfy the equal-time ``dual'' order and disorder field algebra,
$$
\opsigma (x^0 , x^1 ; \beta)\,\opmu (x^0 , y^1 ; \beta )\,=
$$
\be
\,\opmu (x^0 , y^1 ;\beta )\,\opsigma (x^0 , x^1 ; \beta) \,
e^{\textstyle\,i\,a\,b\,\gamma^5\,\theta (y^1 - x^1)}\,,
\ee
$$
\widetilde{\opsigma} (x^0 , x^1 ; \beta)\,\widetilde{\opmu} (x^0 , y^1 ; \beta )\,=
$$
\be
\widetilde{\opmu} (x^0 , y^1 ;\beta )\,\widetilde{\opsigma} (x^0 , x^1 ; \beta) \,
e^{\textstyle\,i\,a\,b\,\gamma^5\,\theta (y^1 - x^1)}\,,
\ee

\no where $\theta$ is the step function. The commutation between $\opsigma$ ($\widetilde{\opsigma}$) and $\opmu$ ($\widetilde{\opmu}$) produces a dislocation in the field $\phi (x , \beta)$ ($\widetilde\phi (x ; \beta )$) if $\opsigma$ ($\widetilde{\opsigma}$) is to the right of $\opmu$ ($\widetilde{\opmu}$) and leaves it unchanged otherwise. The operator $\opmu (x ; \beta )$ ($\widetilde{\opmu} ( x ; \beta )$) does not change the field $\widetilde\phi (y ; \beta )$ ($\phi (y ; \beta ))$,
\be
[\opsigma (x ; \beta)\,,\,\widetilde{\opmu} (y ; \beta )]\,=\,0\,\,\,,\forall\,(x , y)\,,
\ee
\be
[\widetilde{\opsigma} (x ; \beta)\,,\,\opmu (y ; \beta )]\,=\,0\,\,\,,\forall\,(x , y)\,.
\ee

As in the $T = 0$ case, the the bosonized Fermi thermofields are local ``dyon'' fields written in terms 
of order and disorder thermal variables,
\be\label{p1}
\psi(x ; \beta ) \,=\, f (\varepsilon )\, \opsigma (x ; \beta )\,\opmu (x ; \beta)\,,
\ee
\be\label{tp2}
\widetilde\psi (x ; \beta) \,=\,f (\varepsilon )\, \widetilde{\opsigma} (x ; \beta )\,\widetilde{\opmu} (x ; \beta)\,.
\ee

The general correlation functions of the thermal operators $\opmu (x ; \beta )$ and $\opsigma (x ; \beta )$ are defined with respect to the Fock vacuum $\vert \widetilde 0 , 0 \rangle\,=\,\vert \widetilde 0 \rangle \otimes \vert 0 \rangle $ and are given in terms of the light-cone coordinates $x^\pm = x^0 \pm x^1$ by
$$
\langle 0 , \widetilde 0 \vert\prod_{i =1}^n \opsigma (x_i ; \beta ) \prod_{j = 1}^n \opmu (y_j ; \beta )
\prod_{k = 1}^n \opsigma^\ast (\bar x_k ; \beta ) \prod_{\ell = 1}^n \opmu^\ast (\bar y_\ell ; \beta )\vert \widetilde 0 , 0 \rangle 
$$
$$
= \prod_{i < i^\prime}^n\,\Big [F (x^-_i - x^-_{i^\prime} ; \beta )\,F (x^+_i - x^+_{i^\prime} ; \beta )\,\Big ]^{\,\textstyle \frac{a^2}{4\pi}\,\gamma^5_{x_i} \gamma^5_{x_{i^\prime}}}\times
$$
$$
\prod_{j < j^\prime}^n\,\Big [F (y^-_j - y^-_{j^\prime} ; \beta )\,F (y^+_j - y^+_{j^\prime} ; \beta )\,\Big ]^{\textstyle\,\frac{b^2}{4 \pi}}\times
$$
$$
\prod_{k < k^\prime}^n\,\Big [F (\bar x^-_k - \bar x^-_{k^\prime} ; \beta )\,F (\bar x^+_k - \bar x^+_{k^\prime} ; \beta )\,\Big ]^{\textstyle\,\frac{a^2}{4 \pi}\,\gamma^5_{\bar x_k} \gamma^5_{\bar x_{k^\prime}}}\times
$$
$$
\prod_{\ell < \ell^\prime}^n\,\Big [F (\bar y^-_\ell - \bar y^-_{\ell^\prime} ; \beta )\,F (\bar y^+_\ell - \bar y^+_{\ell^\prime} ; \beta )\,\Big ]^{\textstyle\,\frac{b^2}{4 \pi}}\times
$$
$$
\prod_{i , k}^n\,\Big [F (x^-_i - \bar x^-_k ; \beta )\,F (x^+_i - \bar x^+_k ; \beta )\,\Big ]^{\textstyle\,-\,
\frac{a^2}{4 \pi}\,\gamma^5_{x_i} \gamma^5_{\bar x_k}}\times
$$
$$
\prod_{j , \ell}^n\,\Big [F (y^-_j - \bar y^-_\ell ; \beta )\,F (y^+_j - \bar y^+_\ell ; \beta )\,\Big ]^{\textstyle\,-\,\frac{b^2}{4 \pi}} \times
$$
$$
\prod_{i , j}^n\,\Big [F (x^-_i -  y^-_j ; \beta ) \Big ]^{\textstyle\,\frac{a b}{4 \pi}\,\gamma^5_{x_i}}\, \Big [\,F (x^+_i - y^+_j ; \beta )\,\Big ]^{\textstyle\,-\,\frac{a b}{4 \pi}\,\gamma^5_{x_i}}\times
$$
$$
\prod_{i , \ell}^n\,\Big [F (x^-_i -  \bar y^-_\ell ; \beta ) \Big ]^{\textstyle\,-\,\frac{a b}{4 \pi}\,\gamma^5_{x_i}}\,\Big [\,F (x^+_i - \bar y^+_\ell ; \beta )\,\Big ]^{\textstyle\,\frac{a b}{4 \pi}\,\gamma^5_{x_i}}\times
$$
$$
\prod_{j , k}^n\,\Big [F (y^-_j -  \bar x^-_k ; \beta ) \Big ]^{\textstyle\,-\,\frac{a b}{4 \pi}\,\gamma^5_{\bar x_k}}\,\Big [\,F (y^+_j - \bar x^+_k ; \beta )\,\Big ]^{\textstyle\,\,\frac{a b}{4 \pi}\,\gamma^5_{\bar x_k}}\times
$$
\be
\prod_{k , \ell}^n\,\Big [F (\bar x^-_k -  \bar y^-_\ell ; \beta ) \Big ]^{\textstyle\,\frac{a b}{4 \pi}\,\gamma^5_{\bar x_k}}\,\Big [\,F (\bar x^+_k - \bar y^+_\ell ; \beta )\,\Big ]^{\textstyle\,-\,\,\frac{a b}{4 \pi}\,\gamma^5_{\bar x_k}},
\ee

\no where the temperature and space-time dependences are given by
\be
F ( z^\pm ; \beta )\,=\,\frac{\beta}{\pi}\,\sinh\,\big [\frac{\pi}{\beta}\,(z^\pm\,-\,i\,\epsilon )\big ]\,.
\ee

\no Formally, in the limit $T \rightarrow 0$ we recover the general order- disorder correlation functions of the standard massless Thirring model.

Using the thermofield bosonization we have shown how the $T = 0$ order and disorder fields in $1 + 1$ dimensions generalize to the case of non-zero temperature. In this way the massless Fermi thermofield can be viewed as a product of thermal order and disorder operators. This streamlines the presentation of Ref. \cite{ABR}.

Let us finally remark that the use of two-dimensional order and disorder thermal operators will certainly be relevant in the generalization for finite temperature of other models where at zero temperature general statistics fields play a role, as for instance the X-Y model. Notice that the order-order thermal correlation function is given by
$$
\langle \opsigma (x ; \beta ) \opsigma^\ast (0 ; \beta) \rangle =
$$
\be\label{ocft}
\Big \{\Big (\frac{\beta}{\pi} \Big )^2\,
\sinh [\frac{\pi}{\beta}\,(x^0 + x^1)] \,\sinh [\frac{\pi}{\beta}\,(x^0 - x^1)]\,\Big \}^{\,-\,\frac{a^2}{4 \pi}}\,.
\ee

\no The low temperature limit of the Euclidian version of the correlation function (\ref{ocft}) is 
\be\label{oocf}
\langle \opsigma (x) \opsigma^\ast (0) \rangle = 
\vert \,x \,\vert^{\textstyle\,- \,\frac{a^2}{2\pi}}\,,
\ee

\no and corresponds to the order-order correlation function of the continuous limit of the low temperature regime of the X-Y model \cite{Marino,XY}. In Ref. \cite{Marino} the Kadanoff and Ceva prescription \cite{KC} has been generalized for the computation of order- disorder variables correlation functions in the Ising model for continuous quantum field theories with $U (1)$ symmetry. Since the low temperature limit of the Euclidian version of the order-order thermal correlation function (\ref{ocft}) is the continuous limit of the X-Y model given by (\ref{oocf}), it should be very interesting to investigate whether one can relate the Euclidian version of the correlation function (\ref{ocft}) with the continuous limit of the order-order correlator of the X-Y model for other temperature domain. This question and other related problems are the subject of our present investigation.

{\bf Acknowledgment}: The author is greateful to Brazilian Research Council (CNPq) for partial financial support.

}

\end{document}